\documentclass{emulateapj}
\usepackage{apjfonts}

\newcommand{\msun}{\ensuremath{\mathrm{M}_{\sun}}}
\newcommand{\msunperyr}{\ensuremath{\mathrm{M}_{\sun}\,\mbox{yr}^{-1}}}

\newcommand{\figref}[1]{Fig.~\ref{#1}}
\newcommand{\diskmod}[2]{$(#1,10^{-#2})$}

\begin{document}
\slugcomment{Submitted to ApJ May 12, 2008; revised June 18, 2008}
\title{Disk Truncation and Planet Formation in $\gamma$ Cephei}
\shorttitle{PLANET FORMATION IN $\gamma$ Cep}
\author{H.~Jang-Condell\altaffilmark{1,2,3}, M.~Mugrauer\altaffilmark{4} and T.~Schmidt\altaffilmark{4}}
\altaffiltext{1}{Department of Astronomy, University of Maryland, College Park, MD 20742-2421}
\altaffiltext{2}{Exoplanets and Stellar Astrophysics Laboratory, 
NASA Goddard Space Flight Center}
\altaffiltext{3}{email: hannah@astro.umd.edu}
\altaffiltext{4}{Astrophysikalisches Institut und Universit\"{a}ts-Sternwarte, 
Universit\"{a}t Jena, Schillerg\"{a}\ss{}chen 2-3, 07745 Jena, Germany}
\shortauthors{JANG-CONDELL, MUGRAUER \& SCHMIDT}

\begin{abstract}
The $\gamma$ Cephei system 
is one of the most closely bound binary planet hosts known to date. 
The companion ($\gamma$ Cep B) to the 
planet-hosting star ($\gamma$ Cep A) should have truncated any 
protoplanetary disk around $\gamma$ Cep A, possibly limiting 
planet formation in the disk.  We explore this problem by calculating 
the truncation radii of protoplanetary disk models around 
$\gamma$ Cep A to determine whether or not there is sufficient 
material remaining in the disk to form a planet.  
We vary the accretion rate and viscosity parameter of the disk 
models to cover a range of reasonable possibilities for 
the disks properties and determine that for accretion 
rates of $\geq 10^{-7}\,\msunperyr$ and low viscosity parameter, 
sufficient material in gas and solids exist for planet formation 
via core accretion to be possible.  Disk instability 
is less favored, as this can only occur in the most massive 
disk model with an extremely high accretion rate.  
\end{abstract}


\section{Introduction}

Over the last decade, several stellar and substellar companions of 
exoplanet host stars have been found, mostly in seeing-limited 
wide field imaging surveys
\citep[e.g.,][]{2007MugrauerSeifahrtNeuhauser,2006Raghavan_etal,
2006Chauvin_etal}.
The projected separations are a few tens to several thousand AU, 
with companion masses $\sim0.08 - 1.1\,M_{\odot}$.  
\citet{2007EggenbergerUdry} suggest that exoplanets may be less common 
in binaries closer than 120\,AU\@.  
This apparent lack of close companions to exoplanet host stars 
may indicate that planet formation is hampered by the
gravitational influence of a close massive companion. 
However, further investigations are needed
to confirm this result.

One of the closest planet host binaries presently known is $\gamma$\,Cep.  
The planet orbits the primary $\gamma$\,CepA
on a 906 day orbit and has $m\sin(i)\sim1.7$\,$M_{Jup}$
\citep{1988Campbell_etal,2003Hatzes_etal}.  The stellar companion 
orbits at a semi-major axis of 20 AU with an eccentricity of 0.4
\citep{ 2003Hatzes_etal,2007Torres_gcep}.  Given that massive
companions can disrupt the protoplanetary disks in which planets form,
how feasible is the {\em in situ} formation of a planet around
$\gamma$ Cep A?

\citet{2006Haghighipour_gcep} considered the dynamical stability of 
$\gamma$ Cep in order to put limits on possibility of the existence 
of another planet in the system. 
\citet{2004Thebault_gcep} considered the problem of planet formation via 
core accretion in this system using N-body simulations, assuming 
a density profile consistent with 
the Minimum Mass Solar Nebula (MMSN) model of \citet{hayashi}, 
which is steeper than that produced by viscous accretion disk models.  
\citet{2008PaardekooperThebaultMellema} revisited the 
problem including gas drag and determined that giant 
planet formation by core accretion is feasible in $\gamma$ Cep, although 
they did not address the evolution of the gas in the disk or the 
truncation radius of the disk in detail.  
\citet{2008KleyNelson} assumed a specific disk model and examined 
the fate of planet cores inserted into the disk.  
Rather than do a detailed hydrodynamic simulation of a planet embedded
in the disk, our objectives are to model the protoplanetary disk in
$\gamma$ Cep as a viscous accretion disk and to explore which disk
parameters allow planet formation to occur given that the disk is 
truncated by the stellar companion.

This analysis is similar to 
that done in \citet{HJChd188753} (henceforth Paper 1) for 
the extremely close triple system HD 188753.  
Paper 1 concluded that HD 188753 
was unlikely to support a disk sufficiently massive to 
support planet formation.  Indeed, 
the initial claim of a Jupiter-mass planet in HD 188753 
\citep{HD188753} has since been refuted \citep{2007Eggenberger_etal}.
This does not rule out the possibility that the planet could form 
around a single star or in a wide binary and then undergo 
dynamical evolution, such as through close encounters with 
another star \citep{2005PortegiesZwart_McMillan,2005Pfahl}, 
but this is outside the scope of this paper.  

\section{Model Description}

We adopt orbital parameters for the $\gamma$ Cep system 
from \citet{2007Neuhauser_gcep}, as follows:
primary mass $1.40\,M_{\sun}$, 
secondary mass $0.409\,M_{\sun}$,
eccentricity $0.41$, and 
semi-major axis $20.18$ AU.
We ignore the orbit of the planet, 
since we are interested in pre-planetary conditions of the 
disk around the 1.4 $M_{\sun}$ primary.  
We assume that the stars have not undergone significant 
mass loss or orbital evolution since their formation, 
so we can model the properties for $\gamma$\,CepA based on a
pre-main sequence stellar model for a 1.4 $M_{\sun}$ star.  
Since the typical age of a T Tauri star is 1 Myr, 
we assume this age for our model.

\subsection{Disk Model}

The calculation for the disk models 
is described in detail in Paper 1 and \citet{paper1,paper2}.
We assume an $\alpha$-disk model, 
where the viscosity $\nu$ is given by 
$\nu=\alpha c_s h$ where $c_s$ is the sound speed, $h$ is the 
thermal scale height of the disk, and $\alpha$ is a dimensionless 
parameter \citep{shaksun,pringle}.  The disk temperature is 
set by stellar irradiation at the surface and viscous heating at the 
midplane.  The radial and vertical 
density and temperature structure of the disk 
are calculated iteratively for self-consistency.  
We adopt effective temperature $T_* = 4500$ K, and 
radius $R_* = 3.0\:\mathrm{R}_{\sun}$, 
corresponding to a $M_* = 1.4\,\msun$, 
1 Myr old star with metallicity $Z=0.02$ 
\citep{siess_etal}.  

The two remaining free parameters for our disk models are 
the mass accretion rate onto the star $\dot{M}$, 
and the viscosity parameter $\alpha$.  
The exact values for these parameters are unknown for 
$\gamma$ Cep, and it is likely that these values evolved over time, 
so we explore a range of values for both $\dot{M}$ and $\alpha$ 
to determine which, if any, set of parameters allows for planet 
formation to occur.  
As in Paper 1,
we calculate a grid of disk models, 
with $\alpha\in\{0.001, 0.01,0.1\}$ and 
$\dot{M}\in\{10^{-9},10^{-8},10^{-7},10^{-6},10^{-5},10^{-4}\}\:\msunperyr$.
These parameters are roughly consistent 
with observations of T Tauri stars
\citep[e.g.][]{gullbring,hartmann}, 
including the extremely high and transient accretion rates of FU Ori phenomena 
\citep{2000CalvetHartmannStrom, 1996HartmannKenyon}.
In practice, the models are calculated out to 256 AU, but we 
consider only the material interior to the truncation 
radius to be available for planet formation.  
We refer to a given disk model by the coordinate pair 
$(\alpha,\dot{M})$, so 
that Model \diskmod{0.01}{7} refers to the run with 
$\alpha = 0.01$ and $\dot{M}=10^{-7}\:\msunperyr$.

\subsection{Disk Truncation}

The truncation radii of gaseous 
disks depend on the viscosity and temperature of the gas
\citep[henceforth AL]{1994ArtymowiczLubow}, as opposed 
to planetesimal disks, whose truncation radii can be calculated 
from last stable orbits of test particles \citep{pichardo}.  
The truncation radius of each disk model is calculated following AL, 
as in Paper 1.   
In AL, the truncation radius of a circumstellar disk in a close 
binary is where resonant and viscous torques balance.  
This depends on the mass ratio of the binary ($\mu$), 
the semimajor axis of the orbit ($a$), 
the eccentricity of the orbit ($e$), 
and the Reynolds number of the disk (Re).  
For $\gamma$ Cep, $\mu$, $a$, and $e$ have all been determined 
observationally.  The remaining parameter, Re, 
depends on the structure of the disk, which has long 
since dissipated.  

The Reynolds number is defined as 
\( \mbox{Re} = r v_{\phi} / \nu \)
where $r$ is distance from the star,
$v_{\phi}\equiv\sqrt{GM_*/r}$ is the orbital velocity, 
and $\nu\equiv\alpha c_s h$ is the viscosity of the disk.  
Since $c_s/v_{\phi} = h/r$ and $c_s= \sqrt{kT/\bar{m}}$,  
\begin{equation}
\mbox{Re} = \frac{v_{\phi}^2}{\alpha\,c_s^2} = 
\frac{\bar{m} GM_* }{\alpha k T r }.
\end{equation}
Setting $e=0.41$ for $\gamma$ Cep, we read off truncation radii 
in units of semi-major axis
of the circumprimary disk versus Re from Figs.~5 and 6 in AL, 
for $\mu=0.3$ and $0.1$, respectively.  For $\gamma$ Cep, $\mu=0.22$, 
so we interpolate in $\mu$ to find the final 
truncation radius versus Re relation.  
This relation is plotted as long-dashed black line in 
\figref{reynolds}, for a semi-major axis of 20.18 AU.   

We assume that the disks are dynamically truncated 
and that irradiation from the stellar companion is negligible 
compared to heating from viscous accretion and the central star.  
This irradiation would most likely further 
decrease the likelihood of planet formation since it would provide 
an additional heat source at the outer edge of the disk, 
inhibiting planet formation by either core accretion or disk instability.  
In the absence of additional accretion of material past the companion's 
orbit onto the disk, the disk should be viscously spreading both 
inwards and outwards.  Thus, the calculated truncated disk masses 
should be considered upper limits.

\section{Results}

\begin{figure}[t]
\plotone{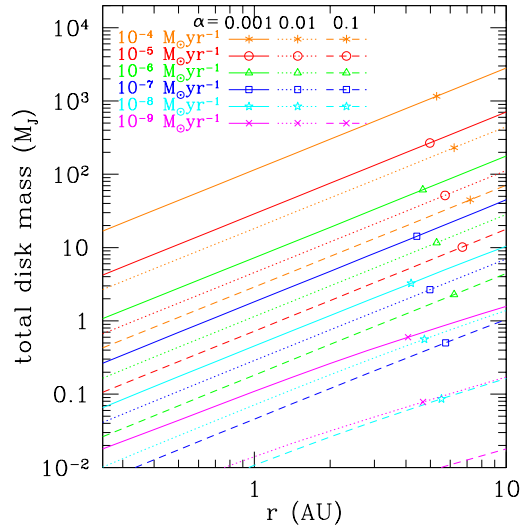}
\caption{\label{mass} 
Enclosed disk mass versus radius for disk models for $\gamma$ Cep, 
in units of $M_J$.
The accretion rate is indicated by color and 
symbol type:
orange asterisks for $10^{-4}$, 
red circles for $10^{-5}$,
green triangles for $10^{-6}$, 
blue squares for $10^{-7}$, 
cyan stars for $10^{-8}$, 
and magenta crosses for $10^{-9}\:\msunperyr$.  
Models with $\alpha$ of 0.001, 0.01 and 0.01 are indicated by 
solid, dotted and dashed lines, respectively.  The locations of the points 
mark the truncation radius and maximum disk mass for each disk model.
}
\end{figure}

In \figref{mass} we plot the mass profiles of the disk models. 
The line colors and types (solid/dotted/dashed) indicate 
$\dot{M}$ and $\alpha$ parameter for each disk model, respectively 
(see legend for details).  Disk mass increases with 
increasing $\dot{M}$ and decreasing $\alpha$.  
The truncation radius for each disk model is indicated by a symbol 
on the line, which also indicates the maximum disk mass for each model.  

\begin{figure}[t]
\plotone{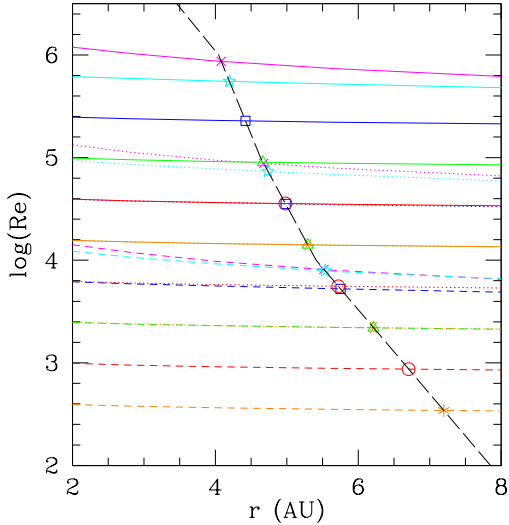}
\caption{\label{reynolds} 
Reynolds number versus radius for each of the disk models, marked 
the same as in \figref{mass}.  The black long-dashed line shows the 
truncation radius versus Reynolds number relation for the 
$\gamma$ Cep binary.  The truncation radius for each model 
is at the intersection with this line.  }
\end{figure}

\figref{reynolds}
shows Re versus radius for each disk model, coded the same 
as in \figref{mass}.  
The Reynolds number in each disk model depends on the input parameters, 
but stays fairly flat with radius.
The black long-dashed line shows the truncation radius versus 
Re relation calculated following AL\@.  
From the intersection of the long-dashed line with each model profile, 
we determine a unique truncation radius for each disk model, 
marked by symbols.

The truncation radii are in the range of 4-7 AU, roughly consistent with 
the 4-5 AU truncation radius assumed by \citet{2004Thebault_gcep}, 
albeit a bit larger.  This is expected because 
viscous torques of a gaseous disk allow it to extend farther than 
a particle-only disk.  
If we consider 1.6 $M_J$, the mass of the planet $\gamma$ Cep Ab, 
to be the minimum mass necessary for in situ planet formation, 
a majority of the disk models satisfy this criterion.  

\section{Discussion: Core Accretion versus Disk Instability}

Having determined that a truncated disk around $\gamma$ Cep A 
can easily contain enough material to form one or several Jupiter 
mass planets, 
we now turn our attention to whether planet formation may take 
place via core accretion or gravitational instability.  

Giant planet formation by core accretion requires sufficient 
mass of solid material to coagulate into a dense core which can then accrete 
gas.  We calculate the mass of dust or solid materials as in 
Paper I, adopting the dust composition from 
\citet{pollack_dust} of olivines, orthopyroxene, iron, water, troilite, 
refractory organics and volatile organics. 
At a given radius in the disk, 
we assume that each species is completely condensed or vaporized, 
depending on whether the temperature is lower or higher,
respectively, than the sublimation temperature for that species.  

\begin{figure}[t]
\plotone{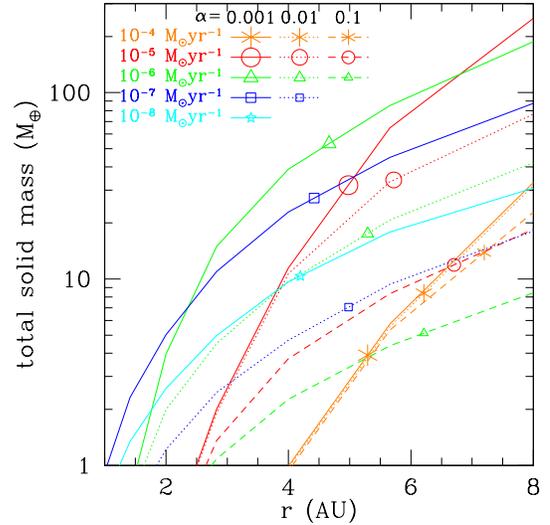}
\caption{\label{solids} Mass of solids for each disk model.  
The position of the point indicates the truncation radius and 
maximum enclosed solid mass.  The sizes of the points indicate the 
relative total disk mass. 
See \figref{mass} for key.}
\end{figure}

In \figref{solids} we plot the mass of solids or dust particles 
in the disk as function of radius.  Here, we only plot those 
disks that contain more than 1.6 $M_J$ of material within 
their truncation radii.  We use 
the same coding of colors, lines and symbols 
to represent the different disks as in \figref{mass}, 
scaling the symbols sizes relative to the disk masses.  
The amount of solids does not simply scale with the mass of the disk 
because hotter disks sublimate their dust.  
While higher accretion rates yield more massive disks, they also 
yield higher temperatures.
On the other hand, higher $\alpha$ values yield less massive disks 
and higher temperatures.  
If we require a minimum of 20 Earth masses (M$_{\earth}$)
of solids to form a giant planet core, we find that 
the Models 
\diskmod{0.001}{6},
\diskmod{0.01}{5},
\diskmod{0.001}{5}, and 
\diskmod{0.001}{7}
all have sufficient solids to form giant planets by core accretion.  

\begin{figure}[b]
\plotone{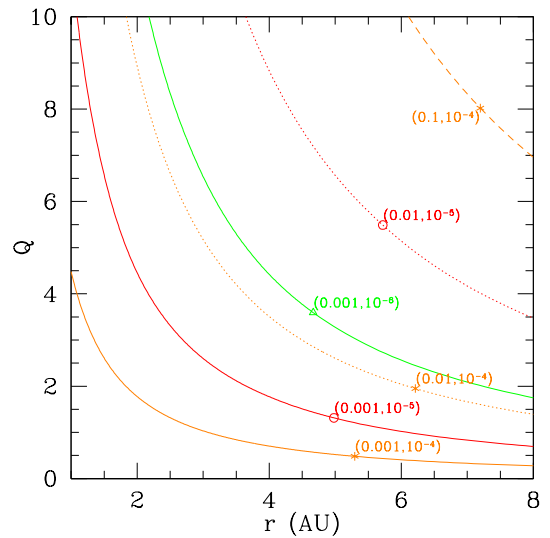}
\caption{\label{Q} Toomre $Q$ parameter for each disk model, as 
indicated by the labels.  The point on each line indicates the 
truncation radius.  Models not shown on this plot have $Q>10$
interior to their truncation radii. }
\end{figure}

The metric for planet formation by gravitational instability 
is the Toomre $Q$ parameter:
\begin{equation}
Q = \frac{c_s \kappa} {\pi G \Sigma}.
\end{equation}
In order for planet formation to proceed, $Q<1$ is required.  
In \figref{Q} we plot the local value of $Q$ versus radius 
for our disk models.  Only six of these have $Q<10$ 
and Model \diskmod{0.001}{4}, the most massive disk, has $Q<1$.
As discussed previously, $10^{-4}\,\msunperyr$ is an extreme 
accretion rate, seen only episodically in FU Ori stars.  
Moreover, $Q<1$ only in very the outermost part of the disk.  
Therefore, we find it unlikely that $\gamma$ Cep Ab
formed through gravitational instability.

These results may be further generalized to suggest that planet 
formation in any close binary is more likely to occur through core 
accretion than disk instability.  This is because disk instability 
happens only in the most massive disks.  While the amount of solids 
is also dependent to some extent on the mass of the disk, the
minimum disk mass for core accretion is still 
below that required for disk instability.  

\section{Conclusions and Future Work}

We assessed the feasibility of {\em in situ} planet formation 
in $\gamma$ Cep by examining the properties of a protoplanetary 
disk around the primary star, given the current orbital parameters 
of the binary star.  We examined a range of accretion rates 
and viscosity parameters and determined the truncation 
radius for each disk model.  We find that $\gamma$ Cep A 
can host a truncated disk of sufficient mass to form a giant planet 
with reasonable accretion rates and viscosity parameters, 
so {\em in situ} planet formation is possible.  
There are sufficient solids in the truncated disk for core 
accretion to occur for accretion rates higher than 
$10^{-7}\,\msunperyr$ and low values of the viscosity 
parameter, $\alpha$.  This is a relatively high accretion rate, 
which may indicate that giant planet formation must take place 
very early on, within $10^5 - 10^6$ years, 
since it appears that accretion 
rates of T Tauri stars decrease with age 
\citep[e.g.][]{2006Sicilia-Aguilar_etal}.  
Giant planet 
formation by disk instability is unlikely to have occured in 
$\gamma$ Cep.  Disk instability also requires an extremely 
high accretion rate, $\sim10^{-4}\,\msunperyr$, a rate 
that is typical of a transient FU Ori phenomenon.  
On the other hand, this may mean that FU Ori outbursts
and giant planet formation by disk instability are correlated. 

We have omitted effects such as shock-heating or 
triggered planet formation by the binary.  Whether these 
effects inhibit \citep[e.g.,][]{2000Nelson} 
or enhance \citep[e.g.,][]{2006Boss_binary} 
planet formation depends on whether cooling times 
are long or short, respectively \citep[see also][]{2005Mayer_etal}.  
Low mass disks that are not self-gravitating are 
not subject to strong shocks, so our results for core accretion 
are unaffected \citep{2005Mayer_etal}.  This 
strengthens our argument that core accretion is the favored 
mechanism for giant planet formation in close binaries.  

The analysis presented here should hold also true for wider binary systems.  
That is, giant planet formation is possible around any star 
where the stellar companion has a mass ratio $\mu\lesssim0.2$ and 
has an orbit wider than $\gamma$ Cep B, with 
disk instability becoming increasingly feasible in wider systems.  
The handful of triple systems with exoplanets that have been 
discovered to date are all hierarchical, 
where the exoplanet host star 
and a close stellar pair revolve around a common barycenter. 
In these cases, the close stellar pair can be treated dynamically as a 
single object.  The closest planet host triple system with confirmed 
exoplanets presently known is HD\,65216\,A+BC, at $\sim$250\,AU 
\citep{2007MugrauerSeifahrtNeuhauser}.  Thus, we 
conclude that there is no barrier to planet formation in the known 
planet-hosting triple systems.  

In \citet{HJChd188753}, the same analysis was carried out for 
a hypothesized planet in HD188753, 
a multiple system which at 
first glance is only slightly closer than $\gamma$ Cep: 
a semi-major axis of 12.3 AU and eccentricity of 0.5. 
{\em In situ} planet formation in HD 188753 was 
ruled out, whereas it is deemed feasible for $\gamma$ Cep.  
Somewhere between the parameters of these 
two systems lies the transition between possiblity and impossibility 
of planet formation in close binaries.  We will explore this parameter 
space and put limits on the closeness of planet host binaries 
in a future paper.  

Another interesting case is a system where the massive companion is 
a white dwarf.  Known white dwarf planet hosts include 
Gl\,86\,B with a projected separation of only 20\,AU; and 
HD\,27442 with a subgiant primary and white dwarf secondary 
\citep{2007MugrauerNeuhauserMazeh}.
Stellar evolution and mass loss generally widens the orbits of binary 
stars, so it is likely that these systems were originally much closer  
\citep[e.g.][]{2002DebesSigurdsson}. 
In order to determine whether 
{\em in situ} planet formation could have taken place in 
these sytems, the original orbital configuration must first be determined 
and the disk truncation radii based on that.  
We will address this issue in a future paper.

\acknowledgements

HJ-C acknowledges support by the NASA Astrobiology Institute under
Cooperative Agreement NNA04CC09A at the Carnegie Institution of Washington, 
Department of Terrestrial Magnetism.  
HJ-C also acknowledges support from a Michelson Fellowship:
this work was performed in part under contract with the Jet
Propulsion Laboratory (JPL) funded by NASA through the Michelson
Fellowship Program. JPL is managed for NASA by the California
Institute of Technology.
TS acknowledges support from Evangelisches Studienwerk e.V.~Villigst.

\bibliographystyle{apj}
\bibliography{apj-jour,../../../planets,../../../jang-condell,../../gammaCep,../../../HD188753/hd188753}

\begin{thebibliography}{36}
\expandafter\ifx\csname natexlab\endcsname\relax\def\natexlab#1{#1}\fi

\bibitem[{{Artymowicz} \& {Lubow}(1994)}]{1994ArtymowiczLubow}
{Artymowicz}, P. \& {Lubow}, S.~H. 1994, \apj, 421, 651

\bibitem[{{Boss}(2006)}]{2006Boss_binary}
{Boss}, A.~P. 2006, \apj, 641, 1148

\bibitem[{{Calvet} {et~al.}(2000){Calvet}, {Hartmann}, \&
  {Strom}}]{2000CalvetHartmannStrom}
{Calvet}, N., {Hartmann}, L., \& {Strom}, S.~E. 2000, Protostars and Planets
  IV, 377

\bibitem[{{Campbell} {et~al.}(1988){Campbell}, {Walker}, \&
  {Yang}}]{1988Campbell_etal}
{Campbell}, B., {Walker}, G.~A.~H., \& {Yang}, S. 1988, \apj, 331, 902

\bibitem[{{Chauvin} {et~al.}(2006){Chauvin}, {Lagrange}, {Udry}, {Fusco},
  {Galland}, {Naef}, {Beuzit}, \& {Mayor}}]{2006Chauvin_etal}
{Chauvin}, G., {Lagrange}, A.-M., {Udry}, S., {Fusco}, T., {Galland}, F.,
  {Naef}, D., {Beuzit}, J.-L., \& {Mayor}, M. 2006, \aap, 456, 1165

\bibitem[{{Debes} \& {Sigurdsson}(2002)}]{2002DebesSigurdsson}
{Debes}, J.~H. \& {Sigurdsson}, S. 2002, \apj, 572, 556

\bibitem[{{Eggenberger} \& {Udry}(2007)}]{2007EggenbergerUdry}
{Eggenberger}, A. \& {Udry}, S. 2007, ArXiv e-prints, 705

\bibitem[{{Eggenberger} {et~al.}(2007){Eggenberger}, {Udry}, {Mazeh}, {Segal},
  \& {Mayor}}]{2007Eggenberger_etal}
{Eggenberger}, A., {Udry}, S., {Mazeh}, T., {Segal}, Y., \& {Mayor}, M. 2007,
  \aap, 466, 1179

\bibitem[{{Gullbring} {et~al.}(1998){Gullbring}, {Hartmann}, {Briceno}, \&
  {Calvet}}]{gullbring}
{Gullbring}, E., {Hartmann}, L., {Briceno}, C., \& {Calvet}, N. 1998, \apj,
  492, 323

\bibitem[{{Haghighipour}(2006)}]{2006Haghighipour_gcep}
{Haghighipour}, N. 2006, \apj, 644, 543

\bibitem[{{Hartmann} {et~al.}(1998){Hartmann}, {Calvet}, {Gullbring}, \&
  {D'Alessio}}]{hartmann}
{Hartmann}, L., {Calvet}, N., {Gullbring}, E., \& {D'Alessio}, P. 1998, \apj,
  495, 385

\bibitem[{{Hartmann} \& {Kenyon}(1996)}]{1996HartmannKenyon}
{Hartmann}, L. \& {Kenyon}, S.~J. 1996, \araa, 34, 207

\bibitem[{{Hatzes} {et~al.}(2003){Hatzes}, {Cochran}, {Endl}, {McArthur},
  {Paulson}, {Walker}, {Campbell}, \& {Yang}}]{2003Hatzes_etal}
{Hatzes}, A.~P., {Cochran}, W.~D., {Endl}, M., {McArthur}, B., {Paulson},
  D.~B., {Walker}, G.~A.~H., {Campbell}, B., \& {Yang}, S. 2003, \apj, 599,
  1383

\bibitem[{{Hayashi}(1981)}]{hayashi}
{Hayashi}, C. 1981, Progress of Theoretical Physics Supplement, 70, 35

\bibitem[{{Jang-Condell}(2007)}]{HJChd188753}
{Jang-Condell}, H. 2007, \apj, 654, 641

\bibitem[{{Jang-Condell} \& {Sasselov}(2003)}]{paper1}
{Jang-Condell}, H. \& {Sasselov}, D.~D. 2003, \apj, 593, 1116

\bibitem[{{Jang-Condell} \& {Sasselov}(2004)}]{paper2}
---. 2004, \apj, 608, 497

\bibitem[{{Kley} \& {Nelson}(2008)}]{2008KleyNelson}
{Kley}, W. \& {Nelson}, R. 2008, ArXiv e-prints, 805

\bibitem[{{Konacki}(2005)}]{HD188753}
{Konacki}, M. 2005, Nature, 436, 230

\bibitem[{{Mayer} {et~al.}(2005){Mayer}, {Wadsley}, {Quinn}, \&
  {Stadel}}]{2005Mayer_etal}
{Mayer}, L., {Wadsley}, J., {Quinn}, T., \& {Stadel}, J. 2005, \mnras, 363, 641

\bibitem[{{Mugrauer} {et~al.}(2007{\natexlab{a}}){Mugrauer}, {Neuh{\"a}user},
  \& {Mazeh}}]{2007MugrauerNeuhauserMazeh}
{Mugrauer}, M., {Neuh{\"a}user}, R., \& {Mazeh}, T. 2007{\natexlab{a}}, \aap,
  469, 755

\bibitem[{{Mugrauer} {et~al.}(2007{\natexlab{b}}){Mugrauer}, {Seifahrt}, \&
  {Neuh{\"a}user}}]{2007MugrauerSeifahrtNeuhauser}
{Mugrauer}, M., {Seifahrt}, A., \& {Neuh{\"a}user}, R. 2007{\natexlab{b}},
  \mnras, 378, 1328

\bibitem[{{Nelson}(2000)}]{2000Nelson}
{Nelson}, A.~F. 2000, \apjl, 537, L65

\bibitem[{{Neuh{\"a}user} {et~al.}(2007){Neuh{\"a}user}, {Mugrauer},
  {Fukagawa}, {Torres}, \& {Schmidt}}]{2007Neuhauser_gcep}
{Neuh{\"a}user}, R., {Mugrauer}, M., {Fukagawa}, M., {Torres}, G., \&
  {Schmidt}, T. 2007, \aap, 462, 777

\bibitem[{{Paardekooper} {et~al.}(2008){Paardekooper}, {Th{\'e}bault}, \&
  {Mellema}}]{2008PaardekooperThebaultMellema}
{Paardekooper}, S.-J., {Th{\'e}bault}, P., \& {Mellema}, G. 2008, \mnras, 386,
  973

\bibitem[{{Pfahl}(2005)}]{2005Pfahl}
{Pfahl}, E. 2005, \apjl, 635, L89

\bibitem[{{Pichardo} {et~al.}(2005){Pichardo}, {Sparke}, \&
  {Aguilar}}]{pichardo}
{Pichardo}, B., {Sparke}, L.~S., \& {Aguilar}, L.~A. 2005, \mnras, 359, 521

\bibitem[{{Pollack} {et~al.}(1994){Pollack}, {Hollenbach}, {Beckwith},
  {Simonelli}, {Roush}, \& {Fong}}]{pollack_dust}
{Pollack}, J.~B., {Hollenbach}, D., {Beckwith}, S., {Simonelli}, D.~P.,
  {Roush}, T., \& {Fong}, W. 1994, \apj, 421, 615

\bibitem[{{Portegies Zwart} \& {McMillan}(2005)}]{2005PortegiesZwart_McMillan}
{Portegies Zwart}, S.~F. \& {McMillan}, S.~L.~W. 2005, \apjl, 633, L141

\bibitem[{{Pringle}(1981)}]{pringle}
{Pringle}, J.~E. 1981, \araa, 19, 137

\bibitem[{{Raghavan} {et~al.}(2006){Raghavan}, {Henry}, {Mason}, {Subasavage},
  {Jao}, {Beaulieu}, \& {Hambly}}]{2006Raghavan_etal}
{Raghavan}, D., {Henry}, T.~J., {Mason}, B.~D., {Subasavage}, J.~P., {Jao},
  W.-C., {Beaulieu}, T.~D., \& {Hambly}, N.~C. 2006, \apj, 646, 523

\bibitem[{{Shakura} \& {Sunyaev}(1973)}]{shaksun}
{Shakura}, N.~I. \& {Sunyaev}, R.~A. 1973, \aap, 24, 337

\bibitem[{{Sicilia-Aguilar} {et~al.}(2006){Sicilia-Aguilar}, {Hartmann},
  {F{\"u}r{\'e}sz}, {Henning}, {Dullemond}, \&
  {Brandner}}]{2006Sicilia-Aguilar_etal}
{Sicilia-Aguilar}, A., {Hartmann}, L.~W., {F{\"u}r{\'e}sz}, G., {Henning}, T.,
  {Dullemond}, C., \& {Brandner}, W. 2006, \aj, 132, 2135

\bibitem[{{Siess} {et~al.}(2000){Siess}, {Dufour}, \& {Forestini}}]{siess_etal}
{Siess}, L., {Dufour}, E., \& {Forestini}, M. 2000, \aap, 358, 593

\bibitem[{{Th{\'e}bault} {et~al.}(2004){Th{\'e}bault}, {Marzari}, {Scholl},
  {Turrini}, \& {Barbieri}}]{2004Thebault_gcep}
{Th{\'e}bault}, P., {Marzari}, F., {Scholl}, H., {Turrini}, D., \& {Barbieri},
  M. 2004, \aap, 427, 1097

\bibitem[{{Torres}(2007)}]{2007Torres_gcep}
{Torres}, G. 2007, \apj, 654, 1095

\end{thebibliography}

\end{document}